\documentclass{appolb}
\usepackage{graphicx}
 \newcommand\beq{\begin{equation}}
 \newcommand\eeq{\end{equation}}
 \newcommand\beqn{\begin{eqnarray}}
 \newcommand\eeqn{\end{eqnarray}}
\def\BA{\begin{eqnarray}}
\def\BE{\begin{equation}}
\def\EA{\end{eqnarray}}
\def\EE{\end{equation}}
\def\beq{\begin{equation}}
\def\eeq{\end{equation}}

\def\beqy{\begin{eqnarray}}
\def\eeqy{\end{eqnarray}}
 \newcommand{\la}{\langle}
 \newcommand{\ra}{\rangle}

\def\fm{\,\mbox{fm}}
\def\GeV{\,\mbox{GeV}}

\def\la{\langle}
\def\ra{\rangle}
\def\BF{\begin{figure}[htb]}
\def\BT{\begin{table}[htb]}
\def\EF{\end{figure}}
\def\ET{\end{table}}

\def\e0{\varepsilon_0}

\def\lsim{\mathrel{\rlap{\lower4pt\hbox{\hskip1pt$\sim$}}
    \raise1pt\hbox{$<$}}}         %less than or approx. symbol
\def\gsim{\mathrel{\rlap{\lower4pt\hbox{\hskip1pt$\sim$}}
    \raise1pt\hbox{$>$}}}         %greater than or approx. symbol
\def\Jpsi{J\!/\!\psi}
\def\psip{\psi^{\,\prime}} 

\def\Y{\Upsilon}
\def\Yp{\Upsilon^{\,\prime}}

\begin{document}
%%%%%%%%%%%%%%%%%%%%%%%%


%----------------------------------------------
\title{
\vspace*{-4.8cm}
%=============================================================
Nuclear effects in coherent photoproduction of heavy quarkonia
%=============================================================
%
%The Paper Title comes Here...%
%----------------------------------------------
\thanks{
Presented at ``Diffraction and Low-$x$ 2022'', Corigliano Calabro (Italy), September 24-30, 2022.
%
%Presented by J. Nemchik at the International
%conference Diffraction and Low-$x$ 2022, 24 - 30 Sept. 2022,
%Corigliano Calabro (Cosenza), Italy
}
\\
}
%----------------------------------------------
\author{
\vspace*{-0.3cm}
J.~Nemchik
\address{
Czech Technical University in Prague,
FNSPE, 
\\
B\v rehov\'a 7,
11519 Prague, Czech Republic \\
Institute of Experimental Physics SAS,
Watsonova 47, 04001 Ko\v sice, Slovakia
\\
~~~~~~~~~~~~~~~~~~~~~~~~~~~~~
\\
}
B.Z.~Kopeliovich
\address{
Departamento de F\'{\i}sica,
Universidad T\'ecnica Federico Santa Mar\'{\i}a,
\\
Avenida Espa\~ na 1680, Valpara\'iso, Chile
}
}
%----------------------------------------------
\maketitle

%----------------------------------------------
\begin{abstract}
\vspace*{-0.40cm}
Coherent photoproduction of heavy quarkonia on nuclear targets is studied within the QCD color
dipole formalism including several main phenomena:
i) The correlation between impact parameter of a collision $\vec b$ and dipole orientation $\vec r$;
ii) The higher-twist nuclear shadowing related to the $\bar QQ$ Fock state of the photon;
iii) The leading-twist gluon shadowing corresponding to higher Fock components of the photon
containing gluons;
iv) Reduced effects of quantum coherence in a popular Balitsky-Kovchegov equation compared to
calculations, which are frequently presented in the literature.
Our calculations of differential cross sections are in good agreement with recent ALICE data on 
charmonium production in ultra-peripheral nuclear collisions. We present also predictions for coherent
photoproduction of other quarkonium states ($\psip$(2S), $\Y$(1S) and $\Yp$(2S)) that can be verified by
future measurements at the LHC.
\end{abstract}
%-----------------------------------------------
\vspace*{-0.2cm}
\PACS{
14.40.Pq,13.60.Le,13.60.-r
}

%
%
%
%\vspace*{-0.2cm}
%%%%%%%%%%%%%%%%%%%%%%
%\section{
%Introduction
%\label{intro}
%}
%%%%%%%%%%%%%%%%%%%%%%
%
%
%

%
%
%
%\vspace*{-0.5cm}
%%%%%%%%%%%%%%%%%%%%%%%%%%%%%%%%%%%%%%%%%%%%%%%%%%%%%%
\section{
Significance of $\vec b-\vec r$ correlation
\label{eloss-v}
}
%%%%%%%%%%%%%%%%%%%%%%%%%%%%%%%%%%%%%%%%%%%%%%%%%%%%%%
%
%
%

The dipole-nucleon electroproduction amplitude within
the {\it color dipole formalism} has the following
factorized form
\cite{Kopeliovich:2021dgx},
\vspace*{-0.3cm}
%========================================================
\BA
\!\!\!\!
\mathcal{A}^N(x,\vec q\,)
= 2\! 
\int\! d^2b\,e^{i\vec q\cdot\vec b}
\int\! d^2r
\int_0^1\!\! d\alpha
\Psi_{V}^{*}(\vec r,\alpha)
\mathcal{A}^N_{\bar QQ}(\vec r, x, \alpha,\vec b\,)
\Psi_{\gamma^\ast}(\vec r,\alpha,Q^2)\,,
\label{1}
\EA
%========================================================
%
where
$\vec q$ is the transverse component of the momentum transfer,
$\alpha$ is the fractional light-front (LF) momentum
carried by a heavy quark or antiquark of the
$\bar QQ$ Fock component of the photon with the transverse
separation $\vec r$.   

The dipole-proton amplitude in Eq.~(\ref{1}),
$\mathcal{A}^N_{\bar QQ}(\vec r, x, \alpha,\vec b\,)$,
depends also on the impact parameter of collision $\vec b$.  
The LF distribution functions
$\Psi_{\gamma^\ast}(\vec r,\alpha,Q^2)$
and $\Psi_V(\vec r,\alpha)$
correspond to transitions
$\gamma^\ast\to \bar QQ$
 and
$\bar QQ\to V$, respectively.

Higher Fock components containing gluons contribute by default to the
dipole-proton amplitude.
Considering nuclear targets, these components must be taken into account
{\it separately} due to different
{\it coherence effects} in gluon radiation.

The essential feature
of
$\mathcal{A}^N_{\bar QQ}(\vec r, x, \alpha,\vec b\,)$
is the $\vec b-\vec r$
correlation \cite{Kopeliovich:2021dgx},
\vspace*{-0.45cm}
%
%================================================================
\BA 
\mathrm{Im} \mathcal{A}^N_{\bar QQ}(\vec r, x, \alpha,\vec b\,)
=
\frac{\sigma_0}{8\pi \mathcal{B}(x)}\,
\Biggl\{
\exp\left[-\,\frac{\bigl [\vec b+\vec
r(1-\alpha)\bigr ]^2}{2\mathcal{B}(x)}\right]
+
\nonumber\\
%\hspace*{-1.0cm}
\exp\left[-\,\frac{(\vec
b-\vec r\alpha)^2}{2\mathcal{B}(x)}\right]
- \,2\,\exp\Biggl[-\,\frac{r^2}{R_0^2(x)}
-\,\frac{\bigl [\,\vec b+(1/2-\alpha)\vec
r\,\bigr ]^2}{2\mathcal{B}(x)}\Biggr]
\Biggr\}\,,
\label{2}
 \EA
%================================================================
%
where the
interaction vanishes if
$\vec r\bot\vec b$ and
reaches maximal strength if
$\vec r\parallel\vec b$.
%
%In our calculations we adopted
%Golec-Biernat-Wusthoff (GBW) 
%GBW \cite{Golec-Biernat:1998zce} and 
%Bartels-Golec-Biernat-Kowalski (BGBK) 
%BGBK
%\cite{Bartels:2002cj} parametrizations.

From the known amplitude (\ref{1}) one can calculate
the differential cross section
%================================================================
\vspace*{-0.3cm}
\BA
\frac{d\sigma^{\gamma N\to V N}(x,t=-q^2)}{dt}
=
\frac{1}{16\,\pi}\,
\Bigl |
\mathcal{A}^{N}(x,\vec q\,)
\Bigr |^2\,,
\label{3}
\EA
%================================================================
%
where
$x = (M_{V}^2+Q^2)/(W^2+Q^2-m_N^2)$
with the quarkonium mass $M_V$.

The real part of the
$\gamma^* N\to V N$ amplitude and the skewness correction
have been incorporated as described in Ref.~\cite{Kopeliovich:2022jwe}.
%
 %*************************************** FIG 1 ******************************************
%
\vspace*{-0.3cm}
\begin{figure}[h]   
    \includegraphics[width=14.0pc,height=8.9pc]{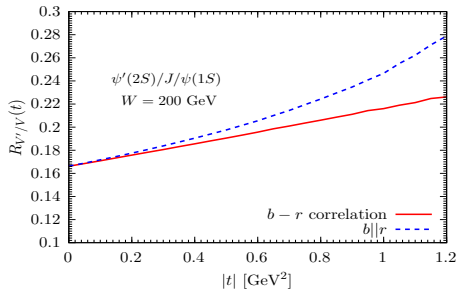}\hspace{1.5pc}
    \begin{minipage}[b]{14.0pc}
    \caption{ \label{fig:b-r}
        Demonstration of the importance of $\vec b-\vec r$ correlation in
        photoproduction of the $\psip(2S)$-to-$\Jpsi(1S)$ ratio 
        of differential cross sections (\ref{3}) at $W=200\,\GeV$.
            }
\vspace*{0.6cm}
    \end{minipage}
\end{figure}
 %****************************************************************************************
%

\vspace*{-0.3cm}
As an example, Fig.~\ref{fig:b-r} nicely demonstrates a manifestation of the $\vec b-\vec r$ correlation
in $\psip(2S)$-to-$\Jpsi(1S)$ ratio of differential cross sections
\cite{Kopeliovich:2021dgx}. Model predictions (solid line) are significantly different
from results based on the standard assumption $\vec b\parallel\vec r$, frequently used in the literature.
However, treating
nuclear targets, the effect of the
$\vec b-\vec r$ correlation is diluted \cite{Kopeliovich:2007fv}.

\vspace*{-0.3cm}
%%%%%%%%%%%%%%%%%%%%%%%%%%%%%%%%%%%%%%%%%%%%%%%%
\section{
Higher-twist nuclear shadowing
\label{ht}
}
%%%%%%%%%%%%%%%%%%%%%%%%%%%%%%%%%%%%%%%%%%%%%%%%
%
%
%

The lowest Fock component of the projectile photon
$|\bar QQ\ra$ has a
small transverse dipole size
$\propto 1/m_Q$, where $m_Q$ denotes the heavy quark mass.
The corresponding shadowing correction is small as well since
is $\propto 1/m_Q^2$, and so it can
be treated as a {\it higher
twist effect}.

For the amplitude of coherent quarkonium electroproduction
on nuclear targets,
$\gamma^* A\to V A$, one can employ the above expression (\ref{1}) for
$\mathcal{A}^N(x,\vec q\,)$,
but replacing the dipole-nucleon by dipole-nucleus amplitude,
\vspace*{-0.3cm}
%======================================================
\BA
\mathcal{A}^A (x,\vec q\,)
\!=\!
2\!
\int \!\! d^2b_A\,e^{i\vec q\cdot\vec b_A}\!\!\!
\int\!\! d^2r\!
\int_0^1\!\!\! d\alpha
\Psi_{V}^{*}(\vec r,\alpha)
\mathcal{A}^A_{\bar QQ}(\vec r, x, \alpha,\vec b_A)
\Psi_{\gamma^\ast}(\vec r,\alpha,Q^2),
\label{4}
\EA
%=====================================================
where $b_A$ is the nuclear impact parameter.

In ultra-peripheral collisions (UPC) at the LHC, the photon virtuality
$Q^2\sim 0$ and the photon energy in the target
rest frame is sufficiently high, so 
{\it the coherence length} exceeds substantially the nuclear radius $R_A$,
%%%%%%%%%%%%%%%%%%%%%%%%%%%%%%%%%%%%%%%%
\vspace*{-0.3cm}
\BA
l_c^{\bar QQ} = 1/q_L =
\frac{W^2+Q^2-m_N^2}{m_N\,(M_V^2+Q^2)}\Biggl|_{Q^2\sim 0}
\approx
\frac{W^2}{m_N M_V^2}\gg R_A\,.
\nonumber
\EA
%%%%%%%%%%%%%%%%%%%%%%%%%%%%%%%%%%%%%%%%
%
Then fluctuations of the dipole size are frozen due to
Lorentz time dilation and one can rely on
{\it the eikonal form} for the dipole-nucleus partial amplitude at impact parameter
$\vec {b}_A$,
\vspace*{-0.4cm}
%============================================================================
\BA
\mathrm{Im} \mathcal{A}^A_{\bar QQ}(\vec r, x, \alpha,\vec b_A)\biggl |_{l_c^{\bar QQ}\gg R_A}
\!\!\!\!\!\!\!\!\!
=
\!
1\! -\! \Biggl [1\! -\! \frac{1}{A}\,\!\!
\int\!\! d^{2} b\,
\mathrm{Im} \mathcal{A}^N_{\bar QQ}(\vec r, x, \alpha, \vec{b}\,)
T_{A}(\vec{b}_A+\vec{b}\,)
\Biggr ]^A\!\!.
\label{5}
\EA
%=============================================================================
%
Here
$T_A(\vec b_A) = \int_{-\infty}^{\infty} dz\,\rho_A(\vec b_A,z)$
is the nuclear thickness functions
normalized as
$\int d^2 b_A\,T_A(\vec b_A) = A$, and
$\rho_A(\vec b_A,z)$ is the nuclear density.

The corresponding expression for the differential cross section
in the limit  
$l_c^{\bar QQ}\gg R_A$ is analogous to that for proton, Eq.~(\ref{3}),
and reads
\vspace*{-0.3cm}
%=================================================================================
\BA
\frac{d\sigma^{\gamma^\ast A\to V A}(x,t=-q^2\,)}{dt}\Biggl |_{l_c^{\bar QQ}\gg R_A}
=
\frac{1}{16\,\pi}\,
\Bigl |
\mathcal{A}^{A}(x,\vec q\,)
\Bigr |^2\,.
\label{6}
\EA
%=================================================================================

%
%
%
\vspace*{-0.6cm}
%%%%%%%%%%%%%%%%%%%%%%%%%%%%%%%%%%%%%%%%
\section{
Leading-twist gluon shadowing
\label{lt}
}
%%%%%%%%%%%%%%%%%%%%%%%%%%%%%%%%%%%%%%%%
%
%
%

The gluon shadowing (GS) corrections are related
to the {\it higher Fock components} of the photon containing, besides the
$\bar QQ$ pair, additional gluons,
$|\bar QQ\,g\ra$,  $|\bar QQ\,2g\ra$, ... $|\bar QQ\,ng\ra$.
In a
$\gamma^*p$ collision these components
are included in the $\bar QQ$-dipole interaction with the proton.
In an electro-production on a nucleus, such multi-gluon fluctuations contribute 
to the amplitude
$\mathcal{A}^N_{\bar QQ}(\vec r, x, \alpha, \vec{b})$ in the eikonal formula
(\ref{5})
for the dipole-nucleus amplitude.

At small photon energies we expect
the {\it Bethe-Heitler regime} of radiation, when each of multiple interactions
produce
{\it independent gluon radiation}.

However, the pattern of multiple interactions changes in the regime of long
$l_c^{\bar QQg}\gg d$, where
$d\approx 2\fm$ is the mean separation between bound nucleons.
The gluon radiation
length reads   
\vspace*{-0.2cm}
%============================================
\BA
l_c^{\bar QQg}
=
\frac{2 k \alpha_g(1-\alpha_g)}{k_T^2+(1-\alpha_g)m_g^2+\alpha_g M_{\bar QQ}^2},
\label{7}
\EA  
%============================================
where
$k$ is the photon energy
in the target rest frame,
$\alpha_g$ is the LF fraction of the photon momentum carried by the
gluon, $M_{\bar QQ}$ is the effective mass of the
${\bar QQ}$ pair and
$m_g\approx0.7\GeV$ is the {\it effective gluon mass} fixed by data on
gluon radiation
\cite{Kopeliovich:1999am,Kopeliovich:2007pq}.
Such a rather large $m_g$ leads to 
a strong inequality
$l_c^{\bar QQg}\ll l_c^{\bar QQ}$, namely
$l_c^{\bar QQg} = l_c^{\bar QQ}/f_g$, where  
a large factor $f_g\approx 10$ has been obtained in
\cite{Kopeliovich:2000ra}.

At long $l_c^{\bar QQg}\gg d$ 
{\it the Landau-Pomeranchuk effect} is at work when
radiation does not resolve multiple interactions acting as one accumulated kick.
This leads to a reduction of
intensity of gluon radiation
compared to the Bethe-Heitler regime.
This is why it is called the GS correction.

Thus the gluon shadowing
is a part of Gribov corrections
corresponding to higher multi-gluon Fock components
of the photon and
requiring eikonalization of these components. 
Differently from $\bar QQ$ fluctuations, a
$\bar QQg$ component {\it does not reach}
the "frozen" size regime due to the
divergent
$d\alpha_g/\alpha_g$ behavior. The corresponding
variation of the
$\bar QQ-g$ dipole size was taken into
account adopting the {\it Green function technique}
\cite{Ivanov:2002kc,Kopeliovich:2022jwe}.

The
$Q\bar Qg$ Fock state is characterized by
{\it two scales}:\\
i) One scale determines
the small
$\bar QQ$ separation, which is
$\approx 1/m_Q$ and represents
{\it a higher twist effect}.
At large $m_Q$
it can be treated as a point-like
color-octet system;
ii) The second scale determines a much larger
$\bar QQ$-$g$ transverse size.
%which is
%independent of
%$m_Q$ (up to Log corrections). \\
\\
Thus
the $\bar QQ-g$ system is {\it strongly asymmetric}
and controls GS, which is the
{\it leading twist effect} since is hardly dependent (only logarithmically) on the
$m_Q$. It can be treated with high precision as {\it glue-glue dipole}
\cite{Kopeliovich:1999am}
with the transverse size 
$\approx 1/m_g$.
The gluon shadowing factor $R_G$ has been calculated as function of $b_A$ and rapidity $y$
adopting the Green function formalism.
The corresponding values of $R_G$ can be found in Fig.1 of Ref.~\cite{Kopeliovich:2022jwe}.

%
%
%
%\vspace*{-0.5cm}
%%%%%%%%%%%%%%%%%%%%%%%%%%%%%%%%%%%%%%%%%%
\section{Comparison with data}\label{data}
%%%%%%%%%%%%%%%%%%%%%%%%%%%%%%%%%%%%%%%%%%
%\vspace*{-0.2cm}
%
%
%

\begin{figure}[h]   
\vspace*{-0.1cm}
    \includegraphics[width=14.2pc]{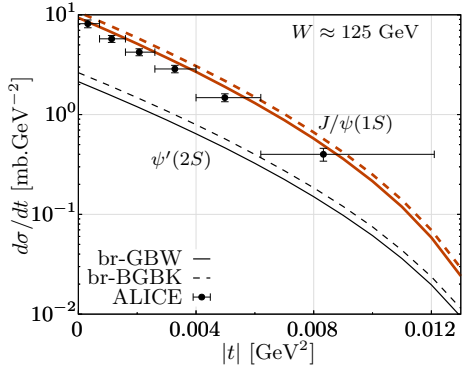}\hspace{1.5pc}
    \begin{minipage}[b]{14.0pc}
    \caption{ \label{fig:data} 
 Predictions for
$d\sigma^{\gamma Pb\to V Pb}/dt$ ~in ~comparison ~with ~ALICE data 
\cite{Acharya:2021bnz}
at 
%c.m. energy
$W\approx 125\,\GeV$.
            }
            \vspace*{0.6cm}
    \end{minipage}
\end{figure}

Model calculations of differential cross sections $d\sigma^{\gamma Pb\to\Jpsi(1S) Pb}/dt$ 
and $d\sigma^{\gamma Pb\to\psip(2S) Pb}/dt$
including effects of $\vec b-\vec r$ correlation, quark and gluon shadowing are presented in Fig.~\ref{fig:data} \cite{Kopeliovich:2022jwe}.
One can see a good agreement of our predictions with ALICE data \cite{Acharya:2021bnz}.
Predictions for other heavy quarkonium states $\Y(1S)$ and $\Yp(2S)$ can be found in \cite{Kopeliovich:2022jwe}.
%
 %*************************************** FIG 2 ******************************************
%
\vspace*{-0.2cm}
%****************************************************************************************
%

%
%
%
\vspace*{-0.5cm}
%%%%%%%%%%%%%%%%%%%%%%%%%%%%%%%%%%%%%%%%%%%%%%%%%%%%%%%
\section{Coherence length for multi-gluon components}
\label{multi}
%%%%%%%%%%%%%%%%%%%%%%%%%%%%%%%%%%%%%%%%%%%%%%%%%%%%%%%
\vspace*{-0.1cm}

The virtual photon with energy $k$ develops a $\bar QQ$ fluctuation for a lifetime
\vspace*{-0.7cm}
%==========================================================================
\beqn
l_c^{\bar QQ} = \frac{2 k}{Q^2+M_{Q\bar Q}^2} = \frac{1}{x m_N} P_q
= l_c^{max} P_q\,
\label{8}
\eeqn
%==========================================================================
where the $\bar QQ$-effective mass squared $M_{\bar QQ}^2 = (m_Q^2 + k_T^2)/\alpha (1-\alpha)$, so that
the factor $P_q = 1/(1+M_{\bar QQ}^2/Q^2)$.
Exact calculations in \cite{Kopeliovich:2000ra} led to mean values
$\la P_{q}\ra_{T,L}\approx 0.36 (0.75)$ at $x\sim 10^{-4}\div 10^{-3}$.
The inequality $\la P_{q}\ra_{L} > \la P_q\ra_T$ means that L photons develop
lighter fluctuations than  T ones.

The higher Fock component
$\bar QQg$ has
the coherence length
\vspace*{-0.3cm}
%================================================
\beqn
l_c^{\bar QQg} = \frac{2 k}{Q^2+M_{\bar QQg}^2} = \frac{1}{x m_N} P_g
= l_c^{max} P_g\,,
\label{9}
\eeqn
%================================================
%
where the $\bar QQg$-effective mass squared
$M_{\bar QQg}^2 =
\frac{M_{\bar QQ}^2 + k_T^2}{1-\alpha_g} +
\frac{m_g^2 + k_T^2}{\alpha_g}
\approx
M_{\bar QQ}^2
(1 + \frac{\gamma}{\alpha_g})$ with
$\gamma = 2 m_g/M_{\bar QQ}^2$, giving thus the
factor
$P_g = \alpha_g/(\alpha_g+\gamma)$.
 Averaging $P_g$
over the gluon radiation spectrum
$d\alpha_g/\alpha_g$
and fixing the
$\bar QQ$ - $g$  
transverse separation at the mean value
$1/m_g$, we obtain
$\la P_g\ra/\la P_q\ra
= 0.12$.

For the $|\bar QQ2g\ra$ Fock state the
effective mass squared
$M_{\bar QQ2g}^2\approx
M_{\bar QQ}^2
(1 + \frac{\gamma}{\alpha_{g1}} + \frac{\gamma}{\alpha_{g2}})$
and the factor
$P_{2g} = \alpha_{g1} \alpha_{g2}/(\alpha_{g1} \alpha_{g2}
+ \gamma\alpha_{g1} + \gamma\alpha_{g2})$.
Performing the averaging process over radiation spectra
$d\alpha_{g1}/\alpha_{g1}$
and
$d\alpha_{g2}/\alpha_{g2}$ we get
$\la P_{2g}\ra/\la P_q\ra
= 0.035$.
Straightforward generalization for higher multi-gluon photon components leads to 
strong inequalities, 
$M_{\bar QQ}^2\ll M_{\bar QQ g}^2\ll 
%M_{Q\bar Q 2g}^2\ll
\cdots\ll M_{\bar QQ ng}^2$
and
$l_c^{\bar QQ}\gg l_c^{\bar QQ g}\gg 
%l_c^{Q\bar Q 2g}\gg
\cdots\gg l_c^{\bar QQ ng}$
with corresponding mean values of $l_c$ at the LHC energy as presented in Tab.~1.

\begin{center}
\vspace{-0.5cm}
%%%%%%%%%%%%%%%%%%%%%%%%%%%%%%%%%%%%%%%%%%%%%%%%%%%%%%%%%%%%%
\BT
\begin{tabular}{|l|c|c|}
 \hline
 & $\la P_{n g}\ra /\la P_q\ra$~
 & $\la l_c\ra$ [fm]
 \\
 \hline
%%%%%%%%%%%%%
 ~$\bar QQ$~
 & -------
 & 120.0
 \\
 \hline
%%%%%%%%%%%%%
 ~$\bar QQg$~
 & 0.11940
 & 14.2
 \\
 \hline
%%%%%%%%%%%%  
 ~$\bar QQ2g$~
 & 0.03560
 & 4.2 
 \\
 \hline
%%%%%%%%%%%
 ~$\bar QQ3g$~
 & 0.01630
 & 1.9
 \\
 \hline   
%%%%%%%%%%%
 ~$\bar QQ4g$~
 & 0.00952
 & 1.1
 \\
 \hline  
%%%%%%%%%%
% ~$Q\bar Q5g$~
% & 0.00639
% & 0.7
% \\
% \hline  
%%%%%%%%%%
% ~$Q\bar Q6g$~
% & 0.00462
% & 0.5
% \\
% \hline  
%%%%%%%%%%
% ~$Q\bar Q7g$~
% & 0.00342
% & 0.4
% \\
% \hline  
%%%%%%%%%%
% ~$Q\bar Q8g$~
% & 0.00256
% & 0.3
% \\
% \hline       
%%%%%%%%%%
\end{tabular}
\hspace*{0.4cm}
%\vspace*{0.5cm}
\begin{minipage}[b]{14.0pc}
%\\
\label{meanP} 
\caption{
\hspace*{0.3cm}
  Fractions of the coherence length for $\bar QQ$ Fock state related to higher
photon components containing different number of gluons.
}  
\vspace*{-1.3cm}
\end{minipage}
\ET
%%%%%%%%%%%%%%%%%%%%%%%%%%%%%%%%%%%%%%%%%%%%%%%%%%%%%%%%%%%% %
\end{center}  

\vspace*{-0.90cm}
In heavy quarkonium production in UPC at the LHC
there are {\it two dominant sources} of shadowing;
the higher twist quark and leading twist gluon shadowing related
to $\bar QQ$ and $\bar QQg$ component of the photon, respectively.
However, the 
%former
quark shadowing 
is suppressed due to large $m_Q$.
Higher Fock states,
$|\bar QQng\ra$ with $n\geq 2$,
%|\bar QQng\ra$, 
have rather small
or negligible contributions to shadowing. 

Balitsky-Kovchegov (BK) equation \cite{balitsky,kovchegov}
in combination with the eikonal expression (\ref{5})
assumes that transverse sizes of
all photon components are 
"frozen" during propagation through the medium,
$l_c^{\bar QQ}$, $l_c^{\bar QQg}$,..., $l_c^{\bar QQng}\gg R_A$.
This leads to exaggeration of shadowing effects.

%
 %*************************************** FIG 3 ****************************************
%
\vspace{-0.3cm}
 \begin{figure}[h]
 \vspace{-0.3cm}
%----------------------------------
    \includegraphics[width=14.0pc,height=8.60pc]{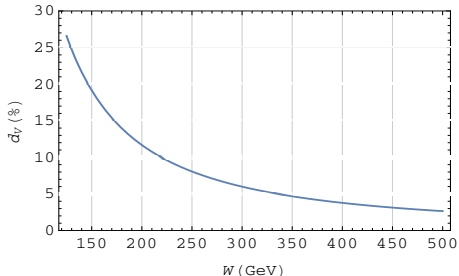}\hspace{1.4pc}
    \begin{minipage}[b]{14.0pc}
    \caption{ \label{fig:coh-bk} 
        Relative impact of reduced coherence effects in the BK equation for
        $\gamma Pb\to\Jpsi Pb$ in terms of the
        energy-dependent factor
        $d_V(W) = |\sigma_{coh}^{GF}-\sigma_{coh}^{eik}|/\sigma_{coh}^{eik}$.
            }
\vspace*{0.60cm}
    \end{minipage}
    \vspace*{-0.6cm}
%----------------------------------
\end{figure}
 %*****************************************************************************************
%

\vspace*{0.30cm}
Fig.~\ref{fig:coh-bk} \cite{prepar} represents a relative comparison of the model predictions for coherent $t$-integrated cross section based on
a solution of BK equation combined with the Green function formalism, $\sigma_{coh}^{GF}$, and with eikonal expression (\ref{5}), $\sigma_{coh}^{eik}$.  One can see that even at the LHC collision energy ($W=125\,\GeV$), the frequently used traditional "eikonal" calculations cause an
{\it overestimation of shadowing effects} by about 20$\%$.
The factor $d_V$ rather slowly decreases with c.m. energy
$W$ and one needs quite large
$W\gsim 500\,\GeV$
in order to use the "frozen" eikonal approximation with a reasonable accuracy.
So one can conclude that the
BK equation cannot be applied to nuclear targets.
\\

\vspace*{-0.30cm}
\textbf{Acknowledgments:}
%=============================================================================
This work was supported in part by ANID-Chile PIA/APOYO AFB180002.
The work of J.N. was partially supported by Grant
No. LTT18002 of the Ministry of Education, Youth and
Sports of the Czech Republic,
by the project of the
European Regional Development Fund No. CZ.02.1.01/0.0/0.0/16\_019/0000778
and by the Slovak Funding Agency, Grant No. 2/0020/22.
%
%The work of M.K. was supported by the project of the International Mobility of Researchers - MSCA IF IV at CTU in Prague 
%CZ.02.2.69/0.0/0.0/20\_079/0017983, Czech Republic.
%==============================================================================

\end{document}